\documentclass[aps,pre,twocolumn,showpacs]{revtex4}

\usepackage{graphicx}
\usepackage{subfigure}
\usepackage{epsfig}
\usepackage{dcolumn}
\usepackage{bm,natbib}
\usepackage{amsmath,amssymb}

\begin{document}

\newcommand{\nwc}{\newcommand}
\nwc{\beq}{\begin{equation}}
\nwc{\eeq}{\end{equation}}
\nwc{\bdm}{\begin{displaymath}}
\nwc{\edm}{\end{displaymath}}
\nwc{\bea}{\begin{eqnarray}}
\nwc{\eea}{\end{eqnarray}}
\nwc{\para}{\paragraph}
\nwc{\vs}{\vspace}
\nwc{\hs}{\hspace}
\nwc{\la}{\langle}
\nwc{\ra}{\rangle}
\nwc{\del}{\partial}
\nwc{\lw}{\linewidth}
\nwc{\nn}{\nonumber}

\nwc{\pd}[2]{\frac{\partial #1}{\partial #2}}
\nwc{\zprl}[3]{Phys. Rev. Lett. ~{\bf #1},~#2~(#3)}
\nwc{\zpre}[3]{Phys. Rev. E ~{\bf #1},~#2~(#3)}
\nwc{\zjsm}[3]{J. Stat. Mech. ~{\bf #1},~#2~(#3)}
\nwc{\zepjb}[3]{Eur. Phys. J. B ~{\bf #1},~#2~(#3)}
\nwc{\zrmp}[3]{Rev. Mod. Phys. ~{\bf #1},~#2~(#3)}
\nwc{\zepl}[3]{Europhys. Lett. ~{\bf #1},~#2~(#3)}
\nwc{\zjsp}[3]{J. Stat. Phys. ~{\bf #1},~#2~(#3)}
\nwc{\zptps}[3]{Prog. Theor. Phys. Suppl. ~{\bf #1},~#2~(#3)}
\nwc{\zpt}[3]{Physics Today ~{\bf #1},~#2~(#3)}
\nwc{\zap}[3]{Adv. Phys. ~{\bf #1},~#2~(#3)}
\nwc{\zjpcm}[3]{J. Phys. Condens. Matter ~{\bf #1},~#2~(#3)}
\nwc{\zjpa}[3]{J. Phys. A ~{\bf #1},~#2~(#3)}




\title{Entropy production theorems and some consequences}

\author{Arnab Saha$^1$, Sourabh Lahiri$^2$ and A. M. Jayannavar$^2$ \email{}}

\date{\today}

\email{jayan@iopb.res.in}

\affiliation{\vspace{0.5cm}$^1$S. N. Bose National Center For Basic Sciences, JD-Block, Sector III, Saltlake, Kolkata -700098, India \\ $^2$ Institute of Phsyics,  Sachivalaya Marg, Bhubaneswar - 751005, India}

\begin{abstract}
The total entropy production fluctuations are studied in some exactly solvable models. For these systems, the detailed fluctuation theorem holds even in the transient state, provided initially the system is prepared in thermal equilibrium. The nature of entropy production during the relaxation of a system to equilibrium is analyzed. The averaged entropy production over a finite time interval gives a higher bound for the average work performed on the system than that obtained from the well known Jarzynski equality. Moreover, the average entropy production as a quantifier for information theoretic nature of irreversibility for finite time nonequilibrium processes is discussed.
\end{abstract}


\maketitle{}

\section{Introduction}

Nonequilibrium thermodynamics of small systems has attracted much interest in recent years \cite{bus05}. In these systems, thermal fluctuations are relevant and probability distributions of physical quantities like work, heat and entropy replace the sharp values of their macroscopic counterparts. 
In this context, fluctuation theorems (FTs) \cite{eva02,har07,rit03,rit06,kur07,eva93,eva94,gal95,jar97,zon02,zon04,nar04,sei05,sei08,cro99} provide exact equalities valid in a system driven out of equilibrium, independent of the nature of driving. One of the fundamental laws of physics, the second law of thermodynamics, states that the entropy of an isolated system always increases. The second law being statistical in nature does not rule out occasional excursions from the typical behaviour. 
FTs make quantitative predictions for observing events that violate the second law within a short time for small systems by comparing the probabilities of entropy generating trajectories to those of entropy annihilating trajectories. FTs  play an important role in allowing us to obtain results generalizing Onsager Reciprocity relations to the nonlinear response coefficients in nonequilibrium state. 

Entropy or entropy production is generally considered as an ensemble property. However, Seifert \cite{sei05,sei08} has generalized the concept of entropy to a single stochastic trajectory. The total entropy production along a single trajectory involves both the particle entropy and the entropy change in the environment. 
It is shown to obey the integral fluctuation theorem (IFT) for any initial condition and drive, over an arbitrary finite time interval, i.e., transient case. In \cite{sei05,sei08}, it is also shown that in the nonequilibrium steady state over a finite time interval, a stronger fluctuation theorem, namely the detailed fluctuation theorem (DFT) holds. Note that originally DFT was found in simulations of two-dimensional sheared fluids \cite{eva93} for entropy production in the medium in the steady state, but in the long-time limit. This was proved in various contexts, e.g. (i) using chaotic hypothesis by Gallavotti and Cohen \cite{gal95}, (ii) using stochastic dynamics by Kurchan \cite{kur98} as well as by  Lebowitz and Spohn \cite{leb99}, and (iii) for Hamiltonian systems by Jarzynski \cite{jar00}. 

In our present work, we obtain the total entropy production ($\Delta s_{tot}$) distribution function, $P(\Delta s_{tot})$, for different classes of solvable models. In particular, we consider
 (i) a Brownian particle in a harmonic trap subjected to an external time-dependent force, and
(ii) a Brownian particle in a harmonic trap, the centre of which is dragged with an arbitrary time-dependent protocol.

In these models, we show that the DFT is valid \emph{even in the transient case}, provided the initial distribution of the state variable is a canonical one. If the initial distribution is other than canonical, DFT in transient case does not hold, as expected. To illustrate this, we have analyzed the total entropy production for a system initially prepared in nonequilibrium state which relaxes to equilibrium. Finally we briefly discuss the important consequences of entropy production fluctuation theorem, namely,
(i) it gives a new bound for the average work done during a nonequilibrium process over a finite time, generalizing the earlier known concept of free energy to a time-dependent nonequilibrium state. This bound is shown to be higher than that obtained from the Jarzynski equality;
(ii) average total entropy production over a finite time quantifies irreversibility in an information theoretic framework via the concept of relative entropy. This is distinct from the recently studied measure \cite{kaw07,gom08a,gom08b,gom07,hor09}.

\section{The model}

\subsection{ Case I: A particle in a harmonic trap subjected to an external time-dependent force}

We consider a Brownian particle in a harmonic potential and in contact with a heat bath at temperature $T$. The system is then subjected to a general driving force $f(t)$. The potential is given by $V_0(x)=\frac{1}{2}kx^2$. The particle dynamics is governed by the Langevin equation in the overdamped limit:

\beq
\gamma \dot{x} = -kx + f(t) + \xi(t),
\label{eq1}
\eeq

where $\gamma$ is the friction coefficient, $k$ is the spring constant and $\xi(t)$ is the Gaussian white noise with the properties $\la \xi(t)\ra=0$ and $\la\xi(t)\xi(t')\ra=2T\gamma\delta(t-t')$. The magnitude of the strength of white noise ensures that the system reaches equilibrium in the absence of time-dependent fields.

Using the method of stochastic energetics (or the energy balance) \cite{sek97,dan02}, the values of physical quantities such as injected work or thermodynamic work ($W$), change in internal energy ($\Delta U$) and heat ($Q$) dissipated to the bath can be calculated for a given stochastic trajectory $x(t)$ over a finite time duration $t$:

\begin{subequations}
\beq
W=\int_0^t\pd{U(x,t')}{t'}dt'=-\int_0^t x(t')\dot{f}(t')dt',
\label{eq2a}
\eeq
\beq
\hspace{1cm}\Delta U = U(x(t),t)-U(x_0,0)= \frac{1}{2}kx^2 - xf(t) - \frac{1}{2}kx_0^2, 
\label{eq2b}
\eeq
and
\beq
 Q=W-\Delta U. 
\label{eq2c}
\eeq
\end{subequations}

Equation (\ref{eq2c}) is a statement of the first law of thermodynamics. The particle trajectory extends from initial time $t=0$ to final time $t$, $x_0$ in equation (\ref{eq2b}) is the initial position of the particle. For simplicity, we have assumed that $f(0)=0$.

Initially the system is prepared in thermal equilibrium. The distribution function is given by

\beq
P(x_0) = \sqrt{\frac{k}{2\pi T}}\exp\left(-\frac{kx_0^2}{2T}\right).
\label{eq3}
\eeq

\noindent The Boltzmann constant $k_B$ has been absorbed in $T$.
The evolved distribution function $P(x,t)$, subjected to the initial condition $P(x_0)$, is obtained by solving the corresponding Fokker Planck equation, and is given by

\beq
P(x,t) = \sqrt{\frac{k}{2\pi T}}\exp\left(-\frac{k(x-\la x\ra)^2}{2T}\right),
\label{eq4}
\eeq
where
\beq
\la x\ra = \frac{1}{\gamma}\int_0^t e^{-k(t-t')/\gamma}f(t')dt'.
\label{eq5}
\eeq

\noindent A change in the medium entropy ($\Delta s_m$) over a time interval is given by

\beq
\Delta s_m = \frac{Q}{T}
\label{eq6}.
\eeq
The nonequilibrium entropy $S$ of the system is defined as 

\beq
S(t) = -\int dx ~P(x,t) \ln P(x,t) = \la s(t)\ra.
\label{eq7}
\eeq
This leads to the definition of a trajectory dependent entropy of the particle as 

\beq
s(t)=-\ln P(x(t),t),
\label{eq8}
\eeq
The change in the system entropy for any trajectory of duration $t$ is given by

\beq
\Delta s = -\ln \left[ \frac{P(x,t)}{P(x_0)} \right],
\label{eq9}
\eeq
where $P(x_0)$ and $P(x,t)$  are the probability densities of the particle positions at initial time $t=0$ and final time $t$ respectively. Thus for a given trajectory $x(t)$, the system entropy $s(t)$ depends on the initial probability density and hence contains the information about the whole ensemble. The total entropy change over time duration $t$ is given by

\beq
\Delta s_{tot} = \Delta s_m + \Delta s.
\label{eq10}
\eeq

Using the above definition of total entropy production, Seifert \cite{sei05,sei08} has derived the IFT , i.e.,

\beq
\la e^{-\Delta s_{tot}}\ra = 1,
\label{eq11}
\eeq
where angular brackets denote average over the statistical ensemble of realizations, or over the ensemble of finite time trajectories.

In nonequilibrium steady state, where the system is characterized by time-idependent stationary distribution, a stronger fluctuation theorem (DFT) valid over arbitrary finite time interval holds \cite{sei05,sei08}. This theorem for the total entropy production can be stated as 

\beq
\frac{P(\Delta s_{tot})}{P(-\Delta s_{tot})} = e^{\Delta s_{tot}}.
\eeq
The above theorem holds even under more general situation, i.e. when system is subjected to periodic driving: $f(x,\tau)=f(x,\tau+\tau_p)$, where $\tau_p$  is the period. The additional requirement is that the system has to settle into a time-periodic state: $P(x,\tau)=P(x,\tau+\tau_p)$, and trajectory length $t$ is an integral multiple of $\tau_p$. 

As a side remark, we would like to state that if the distribution $P(\Delta s_{tot})$ is a Gaussian and satisfies IFT, then it naturally satisfies DFT, even if system is in a transient state. This happens to be the case in our present problem only under the condition that the system is being prepared initially in equilibrium, as shown below.

Using (\ref{eq2c}), (\ref{eq6}), (\ref{eq8}) and (\ref{eq10}), the total entropy becomes

\beq
\Delta s_{tot} = \frac{W-\Delta U}{T}-\ln\frac{P(x,t)}{P(x_0)}
\label{eq13}
\eeq
Substituting for $\Delta U$ from equation (\ref{eq2b}), and using (\ref{eq3}) and (\ref{eq4}), we get

\beq
\Delta s_{tot}= \frac{1}{T}\left(W + \frac{1}{2}k\la x\ra^2 + xf - kx\la x\ra\right),
\label{dstot}
\eeq
The work $W$ is a linear functional of $x(t)$, and from the above equation, we observe that $\Delta s_{tot}$ is linear in $x$, while $x$ is itself a linear functional of Gaussian random variable $\xi(t)$:

\beq
x(t)=x_0e^{-kt/\gamma} + \frac{1}{\gamma}\int_0^t e^{-k(t-t')/\gamma}[f(t')+\xi(t')]dt'.
\label{eqxt}
\eeq
From the above fact it follows that $P(\Delta s_{tot})$ is a Gaussian function. It is therefore sufficient to calculate the mean($\la\Delta s_{tot}\ra$) and  variance ($\sigma^2 \equiv \la\Delta s_{tot}^2\ra - \la\Delta s_{tot}\ra^2$) to get the distribution, which is of the form

\beq
P(\Delta s_{tot}) = \frac{1}{\sqrt{2\pi\sigma^2}}\exp\left(-\frac{(\Delta s_{tot}-\la\Delta s_{tot}\ra)^2}{2\sigma^2}\right)
\eeq
where
 
\beq
\la\Delta s_{tot}\ra = \frac{1}{T}\left(\la W\ra - \frac{1}{2}k\la x\ra^2 + \la x\ra f\right),
\label{eqstotavg}
\eeq

The formal expression of $\la W\ra$ is given by 

\beq
\la W\ra = -\int_0^t\la x(t')\ra \dot{f}(t')~dt 
\eeq
where $\la x\ra$  is given by (\ref{eq5}).
The variance $\sigma^2$ is given by

\begin{subequations}

\bea
\sigma^2 &=& \frac{1}{T}\left(\frac{\la W^2\ra-\la W\ra^2}{T}+\frac{f^2}{k} + k\la x\ra^2 - 2\la x\ra f \right) \nn\\
&& +\frac{1}{T^2}(\la Wx\ra-\la W\ra\la x\ra)(2f-2k\la x\ra) \\
\nn\\
&=& \frac{1}{T}\left(2\la W\ra + \frac{2f^2}{k} + k\la x\ra^2 - 2\la x\ra f \right) \nn\\
&& +\frac{1}{T^2}(\la Wx\ra-\la W\ra\la x\ra)(2f-2k\la x\ra).
\label{eqvar1}
\eea

\end{subequations}





To arrive at (\ref{eqvar1}), we have used the fact that $\la W^2\ra-\la W\ra^2 = 2T\left(\la W\ra+\frac{f^2(t)}{2k}\right)$ which has been proved in appendix \ref{appA}. Also in the same appendix, we have shown that the cross-correlation

\beq
\la Wx\ra - \la W\ra\la x\ra = \frac{T}{k}[k\la x(t)\ra-f(t)].
\label{eqcorr1}
\eeq 

Using equation (\ref{eqcorr1}) in (\ref{eqvar1}), it follows that
\vspace{-0.8cm}

\beq
\sigma^2=2\la\Delta s_{tot}\ra.
\eeq

The Gaussian distribution of $P(\Delta s_{tot})$ along with the above obtained condition for variance implies validity of the detailed fluctuation theorem for general protocol $f(t)$. Needless to say, this theorem in the considered linear system is valid even in the transient case \emph{provided} the initial distribution for the state variable is a canonical distribution. Further, DFT also implies IFT (but the converse is not true).

\begin{figure}
\centering
\vspace{1cm}
\epsfig{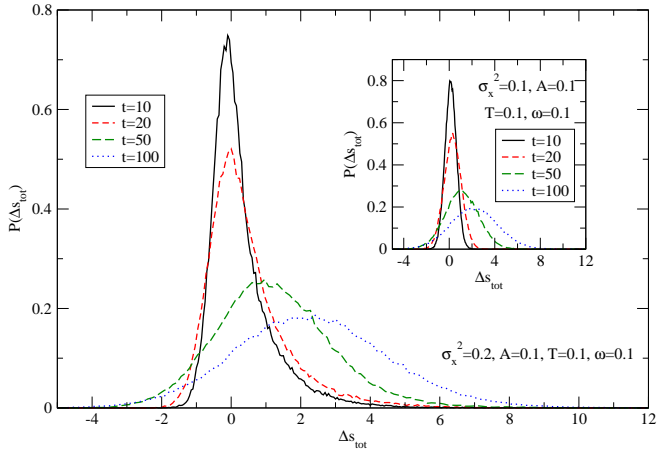}
\caption{(colour online) In the figure, we have plotted $P(\Delta s_{tot})$  vs $\Delta s_{tot}$ for various observation times, when the initial distribution is athermal ($\sigma_x^2=0.2$). For thermal distribution, $\sigma_x^2=0.1$. The observation times are $t=10$ (solid line), $t=20$ (dashed line), $t=50$ (bigger dashed line) and $t=100$ (dotted line). The inset shows total entropy distributions for same observation time values, when the the initial distribution is thermal. For this case, all distributions are Gaussian. For both cases, $A=0.1,~ k=0.1$ and $\omega=0.1$.}
\label{fig1}
\end{figure}

\subsubsection{ Special case: sinusoidal perturbation}

For this case, we consider $f(t)$ to be a sinusoidal oscillating drive, i.e., $f(t)=A\sin\omega t$. Using equation (\ref{eqstotavg}), we obtain

\begin{widetext}
\begin{eqnarray}
\la\Delta s_{tot}\ra &=& \frac{1}{T}\left[\la W\ra-\frac{1}{2}k\la x\ra^2+A\la x\ra\sin\omega t\right] \nn\\\nn\\
&=& \frac{A^2\gamma\omega}{4T(k^2+\gamma^2\omega^2)}\left[2\omega\left\{k^2t+\left(-2-e^{-2kt/\gamma}\right)k\gamma+t\gamma^2\omega^2\right\}\right.\nn\\
&&\left.+8e^{-kt/\gamma}k\gamma\omega\cos\omega t-2k\gamma\omega\cos 2\omega t+(k^2-\gamma^2\omega^2)\sin 2\omega t\right].
\end{eqnarray}
\end{widetext}

The variance is $\sigma^2=2\la\Delta s_{tot}\ra$, and distribution $P(\Delta s_{tot})$ is Gaussian as mentioned earlier.
For this case, if the initial distribution is not canonical, then $P(\Delta s_{tot})$  is not a Gaussian. This is shown in figure \ref{fig1} where we have plotted $P(\Delta s_{tot})$ for the above protocol obtained numerically for various times as mentioned in the figure caption. 
The initial distribution is a Gaussian with $P(x_0)=\sqrt{\frac{k}{2\pi\sigma_x^2}}\exp\left(-\frac{kx_0^2}{2\sigma_x^2}\right)$, where the condition $\sigma_x^2\ne T$ represents an athermal distribution. In the inset, we have plotted $P(\Delta s_{tot})$ for same parameters used for the main figure for thermal initial distribution: $\sigma_x^2=T=0.1$ (for this case, distributions for $\Delta s_{tot}$ are Gaussian). 
All quantities are in dimensionless units and values of physical parameters are mentioned in figure caption. We clearly notice that the distributions $P(\Delta s_{tot})$ in the main figure are non-Gaussian. The observed values of $\la e^{-\Delta s_{tot}}\ra$ from our simulation equal 1.005, 1.006, 0.995 and 1.011 for $t=10, 20, 50$ and 100 respectively in the athermal case. 
All these values are close to unity within our numerical accuracy, clearly validating IFT. For numerical simulations, we have used Heun's scheme. This gives a global error in the dynamics of the order of $h^2$, where $h$ is the time step taken in the simulation (for details, refer to \cite{man00}). To minimize the error in calculating $\la e^{-\Delta s_{tot}}\ra$, we have taken large number of realizations (more than $10^5$), depending on parameters. Our estimated error bars are found to be around $10^{-4}$.  Moreover, these values act as a check on our numerical simulations \cite{sai07,sah08,sin08,lah09}. 
 As the observation time of trajectory increases, weight on the negative side of $P(\Delta s_{tot})$ decreases, i.e., the number of trajectories for which $\Delta s_{tot}<0$ decreases. This is expected as we go to macroscopic scale in time. 
The asymmetric distributions at short time scales tends closer to being a Gaussian distribution with non-zero positive $\la\Delta s_{tot}\ra$. The central Gaussian region increases with the time of observation. The presence of non-Gaussian tails (large deviation functions associated with the probability of extreme events) at large values of $\Delta s_{tot}$ becomes very difficult to detect numerically. However, they are not ruled out. For large times,$\sigma^2 \approx 2\la\Delta s_{tot}\ra$, suggesting validity of DFT only in the time asymptotic regime. Similar observations have been made in regard to work and heat distributions for a driven Brownian particle \cite{sai07,sah08,sin08,lah09,jou06}.

 The Fourier transform of the distribution $P(\Delta s_{tot})$ can be obtained analytically for a given initial athermal Gaussian distribution of the particle position in presence of a drive. This can be obtained following exactly the same procedure of Zon et al \cite{zon04} for heat fluctuations. However, later we consider a simpler case of a system relaxing to equilibrium in absence of protocol (case III).

\subsection{ Case II: $P(\Delta s_{tot})$ for particle in a dragged harmonic oscillator} 

For this case, the effective potential $U(x,t)$ for the Brownian particle is given by

\beq
U(x,t)=\frac{1}{2}k\left(x-\frac{f(t)}{k}\right)^2.
\eeq
The centre of the harmonic oscillator is moved with a time-dependent protocol $f(t)/k$. The special case of this model is when $f(t)/k=ut$ (centre of the oscillator is moved uniformly with velocity $u$). 
This model has been extensively studied both experimentally \cite{col05} and theoretically \cite{zon02,zon04,maz99,spe05,jay08} in regard to analysis of Jarzynski non-equilibrium work relation \cite{jar97} and related issues. 

The expression for work is given by

\beq
W(t)\equiv\int_0^t\pd{U}{t'}dt' = -\int_0^t x(t')\dot{f}(t')dt' + \frac{f^2(t)}{2k}.
\eeq
By taking canonical initial condition for $P(x_0)$, given in equation (\ref{eq3}),
the probability density at time $t$ is given by

\beq
P(x,t) = \sqrt{\frac{k}{2\pi T}}\exp\left(-\frac{k(x-\la x\ra)^2}{2T}\right).
\eeq
where 

\beq
\la x\ra = \frac{1}{\gamma}\int_0^t e^{-k(t-t')/\gamma}f(t')dt'.
\label{eqxavg}
\eeq
The change in internal energy during a time $t$ is 

\beq
\Delta U = \frac{1}{2}k\left(x-\frac{f(t)}{k}\right)^2 - \frac{1}{2}kx_0^2
\eeq
For simplicity, we have set $f(0)=0$.
The expression for $\Delta s_{tot}$ reduces to

\beq
\Delta s_{tot} = \frac{W}{T}-\frac{f^2}{2kT}+\frac{xf}{T}+\frac{k\la x\ra^2}{2T}-\frac{kx\la x\ra}{T}.
\label{eqstot1}
\eeq
From equation (\ref{eqstot1}), it follows that $P(\Delta s_{tot})$ is a Gaussian. Carrying out exactly the similar analysis as before (i.e., for case I), after tedious but straightforward algebra, we finally obtain the expressions for mean and variance:

\beq
\la \Delta s_{tot}\ra = \frac{\la W\ra}{T}-\frac{f^2}{2kT}-\frac{k\la x\ra^2}{2T}+\frac{f\la x\ra}{T} 
\eeq
and 
\beq
\sigma^2 = \frac{2\la W\ra}{T}-\frac{f^2}{kT}-\frac{k\la x\ra^2}{T}+\frac{2f\la x\ra}{T}=2\la\Delta s_{tot}\ra,
\label{eqfd}
\eeq
where $\la W\ra = \int_0^t\la x(t')\ra \dot{f}(t')dt'$, and $\la x\ra$ is given in equation (\ref{eqxavg}). The condition (\ref{eqfd}) along with $P(\Delta s_{tot})$ being Gaussian implies validity of both DFT and IFT for $\Delta s_{tot}$.

\subsubsection{ Special case: The dragging force is linear}

We consider $\frac{f(t)}{k}=ut$, i.e., centre of the harmonic trap is being dragged uniformly with velocity $u$. To obtain $P(\Delta s_{tot})$, we need the expression for $\la\Delta s_{tot}\ra$ only:

\beq
\langle \Delta s_{tot}\rangle = \frac{u^2\gamma t}{T}-\frac{u^2\gamma^2}{2kT}\left(1-e^{-kt/\gamma}\right)\left(3-e^{-kt/\gamma}\right).
\eeq
The above expression can be shown to be positive for all times, as it must be.

\subsection{ Case III: Entropy production with athermal initial condition: a case study for a relaxation dynamics}

In this subsection, we study a system relaxing towards equilibrium.
 If initially the system is prepared in a nonequilibrium state, then in absence of any time-dependent perturbation or protocol, it will relax to a unique equilibrium state. The statistics of total entropy production is analyzed. Our system consists of a Brownian particle in a harmonic oscillator ($V_0(x)=\frac{1}{2}kx^2$) and the temperature of the surrounding medium is $T$. The initial distribution of the particle is taken to be

\beq
P(x_0)=\sqrt{\frac{k}{2\pi\sigma_x^2}}\exp\left(-\frac{kx_0^2}{2\sigma_x^2}\right)
\label{eqPx0}
\eeq
Note that when $\sigma_x^2\ne T$, it
represents athermal initial distribution. Since no protocol is being applied, the thermodynamic work done on the system is identically zero.  As time progresses, the distribution evolves with probability density given by

\beq
P(x,t)=\sqrt{\frac{1}{2\pi\la x^2\ra}}\exp\left(-\frac{x^2}{2\la x^2\ra}\right),
\label{eqPxt}
\eeq
where $\la x^2(t)\ra$ is the variance in $x$ at time $t$, which is equal to $\la x^2(t)\ra = \frac{T}{k}+\frac{\sigma_x^2-T}{k}e^{-2kt/\gamma}$. The distribution $P(x,t)$ relaxes to equilibrium distribution as time $t\to\infty$.
Using equation (\ref{eq13}), (\ref{eqPx0}) and (\ref{eqPxt}), we get

\beq
\Delta s_{tot} = -\frac{\Delta U}{T}-\frac{1}{2}\ln\left(\frac{\sigma_x^2}{k\la x^2\ra}\right)-\left(-\frac{x^2}{2\la x^2\ra}+\frac{kx_0^2}{2\sigma_x^2}\right).\nn
\eeq
Now, considering the fact that $\Delta U=\frac{1}{2}k(x^2-x_0^2),$ we arrive at

\beq
\Delta s_{tot} = \frac{k}{2}\left(\frac{\sigma_x^2-T}{T\sigma_x^2}\right)x_0^2 + \frac{1}{2}\left(\frac{T-k\la x^2\ra}{T\la x^2\ra}\right)x^2 - \frac{1}{2}\ln\left(\frac{\sigma_x^2}{k\la x^2\ra}\right)\nn\\
\label{e6.50}
\eeq
This can be written in a simplified form,

\beq
\Delta s_{tot} = \frac{1}{2}\alpha x_0^2 + \frac{1}{2}\beta x^2 + \kappa ,
\label{eq32}
\eeq
where $\alpha = k\left(\frac{\sigma_x^2-T}{T\sigma_x^2}\right)$; $\beta = \left(\frac{T-k\la x^2\ra}{T\la x^2\ra}\right)$ and $\kappa = - \frac{1}{2}\ln\left(\frac{\sigma_x^2}{k\la x^2\ra}\right)$.

The total entropy production is a quadratic function of $x$ and $x_0$ and hence  $P(\Delta s_{tot})$ is not Gaussian. To obtain $P(\Delta s_{tot})$, we have to know the joint distribution of $x_0$ and $x$, namely $P(x_0,x,t)$ which in our problem can be obtained readily and is given by

\beq
P(x_0,x,t) = \frac{1}{2\pi\sqrt{\det {\bf A}}}\exp[\left({\bf a}-\la {\bf a}\ra\right)^\dagger. {\bf A}^{-1}.\left({\bf a}-\la {\bf a}\ra\right)]
\label{eq33}
\eeq
where

\beq
{\bf a}=\left(\begin{array}{c}
x_0\\x
\end{array}
\right),
\eeq
$x_0$ and $x$ being respectively the initial and final positions of the particle. The matrix ${\bf A}$ is defined through

\begin{eqnarray}
{\bf A} &\equiv& \la ({\bf a}-\la {\bf a}\ra).({\bf a}-\la {\bf a}\ra)^\dagger \ra =\la {\bf a.a}^\dagger\ra\nn\\
&=& \left<\left(\begin{array}{c}
x_0\\x \end{array}\right)
\left(\begin{array}{cc}
x_0 & x \end{array}\right)\right>
= \left(\begin{array}{cc}
\la x_0^2\ra & \la xx_0\ra \\
&\\
\la xx_0\ra & \la x^2\ra \end{array}\right) \nn\\
&=&\left(\begin{array}{ccc}
\frac{\sigma_x^2}{k} && \frac{\sigma_x^2}{k}e^{-kt/\gamma} \\
&\\
\frac{\sigma_x^2}{k}e^{-kt/\gamma} && \frac{T}{k}+\left(\frac{\sigma_x^2-T}{k}\right)e^{-2kt/\gamma}\end{array}\right).
\label{}
\end{eqnarray}
With the help of the distribution given in (\ref{eq33}), one can write, using equation (\ref{eq32}), 

\bea
P(\Delta S_{tot},t) &=& \int_{-\infty}^{\infty}dx ~dx_0P(x_0,x,t)~~\nn\\
&\times&\delta\left[\Delta s_{tot}-\left(\frac{\alpha}{2}x_0^2+\frac{\beta}{2}x^2+\kappa\right)\right].
\eea
The evaluation of $P(\Delta s_{tot})$ is a difficult task. However, the  Fourier transform $\widehat{P}(R,t)\left[\equiv \int e^{iR\Delta s_{tot}}P(\Delta s_{tot})d\Delta s_{tot}\right]$ of $P(\Delta s_{tot})$ can be obtained easily. To this end we can carry out the analysis similar to that for heat distribution in a driven harmonic oscillator by Zon et al \cite{zon04}. Finally we get

\beq
\widehat{P}(R,t) = \frac{e^{iR\kappa}}{\sqrt{\det(I-iR{\bf A.B})}}.
\label{eqPhat}
\eeq
The details of this derivation are given in appendix \ref{appB}.
Substituting $R=i$ in the above equation, and we get $\widehat{P}(R=i,t)=\la e^{-\Delta s_{tot}}\ra = 1$, consistent with the IFT (see appendix B for details). From equation (\ref{eqPhat}), we also note that $\widehat{P}(R,t)\ne \widehat{P}(i-R,t)$, indicating that DFT is not valid for this linear problem in the presence of athermal initial distribution. From above equation, we can also obtain average entropy production given by

\begin{eqnarray}
\la \Delta s_{tot}\ra &=& \left.\frac{1}{i}\pd{}{R}\widehat{P}(R,t)\right|_{R=0}\nn\\
&=& \frac{\sigma_x^2-T}{2T}\left(1-e^{-2kt/\gamma}\right)\nn\\
&& - \frac{1}{2}\ln\left[\frac{\sigma_x^2}{T+e^{-2kt/\gamma}(\sigma_x^2-T)}\right].
\end{eqnarray}
Similarly, higher moments can also be obtained with the use of this characteristic function. One can invert the characteristic function to obtain $P(\Delta s_{tot})$ using integral tables. However, the expression is complicated and unilluminating. From the Fourier transform, it is obvious that $P(\Delta s_{tot})$ is non-Gaussian. 

\begin{figure}
\centering
\epsfig{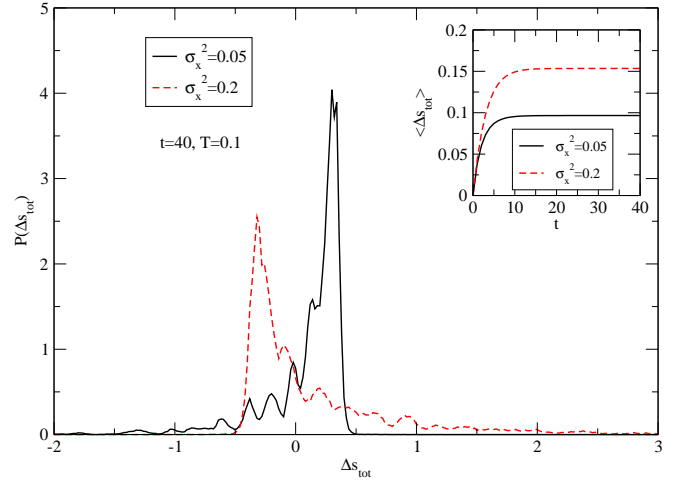}
\caption{(colour online) The figure shows plots of $P(\Delta s_{tot})$ vs $\Delta s_{tot}$ during relaxation to equilibrium (external protocol is absent). The initial distributions are athermal with $\sigma_x^2=0.05$ (solid line) and $\sigma_x^2=0.2$ (dashed line). The spring constant is $k=0.1$ and observation time was $t=40$, by which the system has reached equilibrium (see inset). The inset shows plots average entropy versus observation time for the same initial distributions.}
\label{fig2}
\end{figure}

In figure \ref{fig2}, we have plotted $P(\Delta s_{tot})$ versus $\Delta s_{tot}$ over a fixed time interval (see figure caption) for two different cases for which initial width of the distribution $\sigma_x^2$ equals 0.05 and 0.2. The temperature of the bath is 0.1. The distribution $P(\Delta s_{tot})$ in both cases are asymmetric. 
For the case $\sigma_x^2=0.2$, the distribution is peaked around the negative value of $\Delta s_{tot}$. However, it exhibits a long tail making sure that $\la\Delta s_{tot}\ra$ is always positive. Since initial width of the distribution is larger than the thermal distribution, change in the entropy of the system during the relaxation process is negative and it dominates the total entropy production. 
Hence we obtain peak in $P(\Delta s_{tot})$ in the negative side of $\Delta s_{tot}$. For the case $\sigma_x^2=0.05$, change in the entropy of the system is positive. Hence peak in $P(\Delta s_{tot})$ is in the positive region. In both cases, we obtain $\la e^{-\Delta s_{tot}}\ra$ equal to unity within our numerical accuracy: 0.978 ($\sigma_x^2=0.2$) and 1.001 ($\sigma_x^2=0.05$), consistent with IFT. 
In the inset, we have plotted $\la\Delta s_{tot}\ra$ as a function of time for the above cases. $\la\Delta s_{tot}\ra$ is a monotonically increasing function of time and saturates asymptotically when equilibrium is reached. It may be noted that equilibrium is characterized by zero total entropy production, change in the entropy of bath at any instant being compensated by equal and opposite change in entropy of the system.

\section{Some relations resulting from the average entropy production fluctuations over finite time}

We now discuss some related offshoots of the total entropy production. These give a higher bound for the average work done over a finite time and provide a different quantifier for the footprints of irreversibility. The Jarzynski non-equilibrium work relation \cite{jar97} relates work done over a finite time in a non-equilibrium state to the equilibrium free energy differences, namely,

\beq
\la e^{-\beta W}\ra = e^{-\beta\Delta F}.
\label{eq39}
\eeq
Here the angular brackets denote an average over a statistical ensemble of realizations of a given thermodynamic process. The finite time thermodynamic process involves changing the time dependent parameter $\lambda(t)$ of the system from initial value $\lambda(0)=A$ to a final value $\lambda(\tau)=B$. $\lambda(t)$ can be an arbitrary function of time. Initially the system is prepared in equilibrium state corresponding to parameter $A$, and work $W$ is evaluated over a time $\tau$. At the end of the period $\tau$, the system in general will not be at equilibrium corresponding to parameter $B$, yet from this non-equilibrium work, one can determine the difference in equilibrium free energies, $\Delta F$, between the states described by $A$ and $B$, using equation (\ref{eq39}). From the same equation, using Jensen's inequality, it follows that

\beq
\la W\ra \ge \Delta F = F_B-F_A.
\eeq
This result is consistent with the Clausius inequality, which is written in the form of work and energy, instead of the usual heat and entropy. Using Jensen's inequality and the integral fluctuation theorem of entropy production, namely equation (\ref{eq11}), it follows that the average total entropy production over a time $\tau$, $\la\Delta s_{tot}\ra\ge 0$. Using equation (\ref{eq2c}), this can be rewritten as

\beq
\la\Delta s_{tot}\ra = \frac{1}{T}\la W-\Delta U+T\Delta s\ra \ge 0 \Rightarrow \la W\ra\ge \la\Delta U-T\Delta s\ra,
\label{eq41}
\eeq
where $\Delta U$ and $\Delta s$ are the changes in internal energy and in system entropy respectively. The time-dependent free energy in a nonequilibrium state can be defined as \cite{qia01}:

\beq
F(x,t) = U(x,t)-Ts(x,t) =  U(x,t)+T\ln P(x,t),
\eeq
which is in general a fluctuating quantity. Since free energy depends on entropy, it contains the information of the whole ensemble. In equilibrium, the expectation value of this free energy reduces to the Helmholtz free energy. Using (\ref{eq41}) and the given definition of nonequilibrium free energy described above, it follows that

\beq
\la W\ra \ge \la\Delta F(\tau)\ra,
\label{eqWtau}
\eeq
where $\Delta F(\tau) = F_2(\tau)-F_1(0)$.

If initially the system is prepared in equilibrium with parameter $A$, $F_1$ equals equilibrium free energy $F_A$. $F_2(\tau)$ is determined by the probability distribution at the end point of the protocol at which the system is out of equilibrium with system parameter at $\lambda=B$, i.e. $F_2(\tau)\equiv U(x,\tau)+T\ln P(x,\tau)$. Now in the following, we show that 

\beq
\la\Delta F(\tau)\ra \ge \Delta F = F_B-F_A,
\eeq
thus giving a higher bound for the average work done over a finite time. To this end, consider a situation at which initially the system is prepared in equilibrium with parameter $\lambda=A$ (corresponding to free energy $F_A$) and is allowed to evolve with the time-dependent protocol $\lambda(t)$ up to time $\tau$ at which $\lambda=B$. Beyond $\tau$, the system is allowed to relax to equilibrium by keeping $\lambda$ fixed at B. At the end of the entire process, the total change in equilibrium free energy equals $F_B-F_A$. The free energy being a state function, one can rewrite it as 

\bea
F_B-F_A ~&=&~ \la F_B- F_{2}(\tau) +  F_{2}(\tau) - F_A\ra \nn\\
 ~&=&~  F_B- \la F_{2}(\tau)\ra+\la\Delta F(\tau)\ra.
\label{eqFB-FA}
\eea
Here, $\la\Delta F(\tau)\ra$ is the average change in the nonequilibrium free energy, $\la F_2(\tau)\ra-F_A$, during the process up to time $\tau$ whereas $F_B-\la F_{2}(\tau)\ra$ is the change in the free energy during the relaxation period when the protocol is held fixed. One can readily show that \cite{qia01} during the relaxation process towards equilibrium, the average (or expectation value) of free energy always decreases, i.e., $\la F_B-F_{2}(\tau)\ra$ is negative. From this and equation (\ref{eqFB-FA}), it follows that $\la\Delta F(\tau)\ra \ge F_B-F_A$. Thus we get a higher bound for the average work done than that obtained from the Jarzynski identity \cite{jar97}.

\begin{figure}
\centering
\epsfig{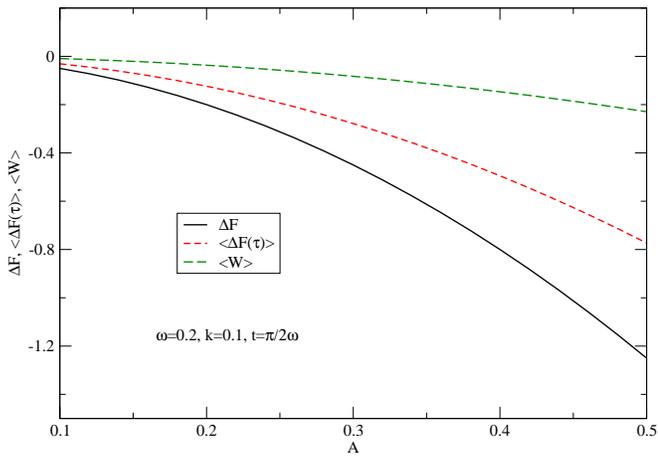}
\caption{(colour online) Plots for $\Delta F, \la\Delta F(\tau)\ra$ and $\la W\ra$ as functions of the driving amplitude $A$, with the parameter values set at $\omega=0.2, k=0.1$ and $\tau=\pi/2\omega$.}
\label{fig3}
\end{figure}

To illustrate this, in figure \ref{fig3} we have plotted $\la W\ra$, $\la\Delta F(\tau)\ra$ and $\Delta F$ for a driven harmonic oscillator $U(x)=\frac{1}{2}kx^2$ with force $f(t)=A\sin\omega t$ as a function of the amplitude of driving $A$. For this graph, system parameter $f(t)$ changes from $f(0)=0$ to $f(\tau)=A$ ~~($\Delta F=F_B-F_A = \frac{-A^2}{2k}$), i.e., for a time variation from $t=0$ to $t=\tau=\frac{\pi}{2\omega}$.
We observe from the figure that $\la\Delta F(\tau)\ra$ is indeed a higher bound. The analytical results for this model are presented in appendix \ref{appC}. 

Some remarks, however, are in order. The realizations for which $W<\Delta F$ need not correspond to $\Delta s_{tot}\le 0$, and vice versa. This implies that the trajectories which violate the second law, namely $\Delta s_{tot}< 0$, do not necessarily violate the inequality $W<\Delta F$, which is also closely related to the second law \cite{jar08}. Equation (\ref{eqWtau}) can be treated as a generalization of Clausius' inequality to nonequilibrium processes.

Dissipation is related to our ability to distinguish the arrow of time. Hence 
the dissipated work $\la W_d\ra=\la W\ra - \Delta F$ is recently identified as the measure of irreversibility. 
Moreover, it turns out that the relative entropy of microscopic trajectories $D_1(P||\widetilde{P})$ in full path space between forward ($P$) and reverse ($\widetilde{P}$) processes is indeed equal to dissipative work, 

\beq
\la W_d\ra = D_1(P||\widetilde{P}).
\eeq
Hence $D_1(P||\widetilde{P})$ works as a measure of irreversibility or indistinguishability between forward and backward evolutions.
 Here, forward evolution corresponds to the system being prepared initially at equilibrium in the state with control parameter $\lambda(0)=A$ evolved up to time $\tau$ at which the control parameter is $\lambda(\tau)=B$.
 During the backward evolution, the system is prepared in equilibrium with control parameter $B$ and the time-reversed protocol is applied from $B$ to $A$. For details, see \cite{kaw07,gom08a,gom08b,gom07}. Separately, it can also be shown by using Crooks identity \cite{gom08a,cro98}.

\beq
\la W_d\ra = D(P(W)||\widetilde{P}(-W))
\label{eqWd}
\eeq
Here $D(P(W)||\widetilde{P}(-W))$ is the relative entropy between the two probability distributions $P(W)$ and $\widetilde{P}(-W)$ which are the work distributions for the same thermodynamic process for forward and backward evolutions respectively. This brings us to an important conclusion that dissipation can be revealed by any finite set of variables which contain information about the work or from the dynamics of those variables which couple to the control parameter $\lambda$. 
Thus one can identify few dynamical variables in which traces of the dissipation reside. 
This is unlike $D(P||\widetilde{P})$, which requires information about entire set of microscopic system variables during their evolution. 

We note that $\la\Delta s_{tot}\ra$ can be taken as the measure of irreversibility as it also represents the relative probability $D_2(P||\widetilde{P})$ between forward and time-reversed backward protocols \cite{sei05,sei08,cro99}:

\begin{widetext}
\begin{equation}
\la\Delta s_{tot}\ra = D_2(P||\widetilde{P}) = \int P(x_0)P[x_\tau|x_0]\ln\left(\frac{P(x_0)P[x_\tau|x_0]}{\widetilde{P}(\tilde{x}_0)\widetilde{P}[\tilde{x}_\tau|\tilde{x}_0]}\right) \mathcal{D}[x_t]~dx_0~dx_\tau.
\end{equation}
\end{widetext}
where $P[x_\tau|x_0]$ and $\widetilde{P}[\tilde{x}_\tau|\tilde{x}_0]$ are the shorthand notations for the probabilities of traversing the entire forward path from $t=0$ to $t=\tau$ described by $x(t)$ and that of traversing the reverse path described by variables $\tilde{x}(\tau-t)$. For details, see references \cite{sei05,sei08,cro99}.
Here, the forward evolution corresponds to the system being prepared initially in any arbitrary state and evolved up to time $\tau$ along a prescribed protocol. At the end of the protocol, the system is in a state $P(x,\tau)$ determined by the initial condition and the dynamics. During the backward process, the system is assumed to be in the \emph{same} state corresponding to the end point of forward evolution $P(x,\tau)\equiv \widetilde{P}(\tilde{x}_0)$ and protocol is time-reversed, thereby evolving the system along the backward trajectory. 
Unlike for work (equation (\ref{eqWd})), there is no Crooks'-like identity for the total entropy production between forward and reverse process (except in the stationary state). Hence it is not possible to describe the measure of irreversibility or dissipation in terms of the relative entropy between probability distribution of $\Delta s_{tot}$ for forward and backward processes. Thus, the information of irreversibility is contained in all the microscopic variables associated with the system. 
This can also be noticed from the fact that the definition of total entropy production, involves the probability density of all the system variables. Moreover, this probability density contains the information about the initial and final ensembles of the system variables.  Identification of $\la\Delta s_{tot}\ra$ as a measure of irreversibility, is tantamount to identifying average dissipative work over a finite time process $\la W-\Delta F(\tau)\ra \equiv \la {W}_d(\tau)\ra$ as a measure of irreversibility, where $\la\Delta F(\tau)\ra$ is the nonequilibrium change in average free energy over a finite time as mentioned before. Needless to say, for this measure $\la W_d(\tau)\ra$, the system need not be in equilibrium at the beginning of the forward process which is a necessary condition for earlier defined measure for irreversibility \cite{kaw07,gom08a,gom08b,gom07}. 
Further work along this direction is in progress.

\section{Conclusions}

In conclusion, we have shown that in a class of solvable linear models, $\Delta s_{tot}$ satisfies DFT even in the transient regime provided the system is initially prepared in an equilibrium state. For athermal initial condition, the nature of total entropy production is analyzed during a relaxation process. 
The bound on average entropy production over a finite time process leads to a higher bound for the average work done over the same finite time interval. Some points have been raised if one assigns meaning to the average entropy production as a measure of irreversibility. 
This measure implies the generalization of Clausius' statement to nonequilibrium finite time processes, namely $\la W_d(\tau)\ra = \la W-\Delta F(\tau)\ra \ge 0$. Analysis of the total entropy production in presence of magnetic field is carried out separately. The results will be published elsewhere \cite{sah09}.

\section{Acknowledgement}

One of us (A.M.J) thanks DST, India for J. C. Bose Fellowship. A.S. thanks IOP, Bhubaneswar (where part of the work is carried out) for hospitality.

\begin{center}  ---------------------  \end{center}


\appendix


\section{Calculation of variance of {\boldmath $\Delta s_{tot}$}}

\label{appA}

\subsection{ Calculation of variance of {\boldmath $W$}:}

\begin{widetext}
Using equation (\ref{eq2a}),

\begin{eqnarray}
W-\la W\ra &=& -\int_0^t (x(t')-\la x(t')\ra)\dot{f}(t')dt' \nn\\
&=& -\int_0^tdt'\dot{f}(t')\left[x(t')e^{-kt'/\gamma}+e^{-kt'/\gamma}\int_0^{t'}e^{kt''/\gamma}\xi(t'')dt''\right]\nn
\end{eqnarray}

\begin{eqnarray}
\therefore \la(W-\la W\ra)^2\ra &=& \la x_0\ra^2\int_0^tdt'\dot{f}(t')e^{-kt'/\gamma}\int_0^tdt_1\dot{f}(t_1)e^{-kt_1/\gamma}\nn\\
&&+\frac{1}{\gamma^2}\int_0^t dt'f(t')e^{-kt'/\gamma}\int_0^tdt_1f(t_1)e^{-kt_1/\gamma}\int_0^{t'}dt''e^{kt''/\gamma}\int_0^{t_1}dt_2e^{kt_2/\gamma}\la\xi(t'')\xi(t_2)\ra \nn\\
&=& \frac{T}{k}\int_0^tdt'\dot{f}(t')e^{-kt'/\gamma}\int_0^tdt_1\dot{f}(t_1)e^{-kt_1/\gamma}\nn\\
&&+\frac{2T}{\gamma}\int_0^t dt'f(t')e^{-kt'/\gamma}\int_0^tdt_1f(t_1)e^{-kt_1/\gamma}\int_0^{t'}dt''e^{2kt''/\gamma}\nn\\
&=& \frac{T}{k}\int_0^tdt_1\dot{f}(t_1)e^{-kt_1/\gamma}\int_0^tdt'\dot{f}(t')e^{kt'/\gamma} \nn\\
&=& \frac{2T}{k}\int_0^tdt_1\dot{f}(t_1)e^{-kt_1/\gamma}\int_0^{t_1}dt'\dot{f}(t')e^{kt'/\gamma}. \nn
\end{eqnarray}

The above integration, when integrated partially gives

$$
\la W^2\ra - \la W\ra^2 = \frac{2T}{k}\int_0^tdt_1\dot{f}(t_1)f(t_1) - \frac{2T}{\gamma}\int_0^tdt_1\dot{f}(t_1)e^{-kt_1/\gamma}\int_0^{t_1}e^{kt'/\gamma}f(t')dt'.
$$

\end{widetext}

Noting that $\la x(t_1)\ra = \frac{e^{-kt_1/\gamma}}{\gamma}\int_0^{t_1}e^{kt'/\gamma}f(t')dt'$ and $W = -\int\la x(t_1)\ra\dot{f}(t_1)dt_1$, we finally get

\bea
\la(W-\la W\ra)^2\ra &=& \frac{2T}{2k}f^2(t) + \frac{2T}{\gamma}\la W\ra \nn\\
&=& 2T\left[\la W\ra+\frac{f^2}{2k}\right].\nn
\eea


\subsection{ Calculation of cross correlation {\boldmath $\la Wx\ra-\la W\ra\la x\ra$}:}
We have, from (\ref{eq2a}) and (\ref{eqxt}),

\begin{widetext}
\begin{eqnarray}
\la W(t)\ra\la x(t)\ra &=& \left[-\int_0^t\la x(t')\ra \dot{f}(t')dt'\right]\times\la x(t)\ra \nn\\
&=& \left[-\int_0^t\left(\frac{1}{\gamma}\int_0^{t'}e^{-k(t'-t'')/\gamma}f(t'')dt''\right)\dot{f}(t')dt'\right]\times \left[\frac{1}{\gamma}\int_0^te^{-k(t-t_1)/\gamma}f(t_1)dt_1\right]\nn\\
&=& -\frac{1}{\gamma^2}\int_0^t dt'\dot{f}(t')\int_0^{t'}dt''e^{-k(t'-t'')/\gamma}f(t'')\int_0^t dt_1 e^{-k(t-t_1)/\gamma}f(t_1).
\label{e11}
\end{eqnarray}

On the other hand,

\begin{eqnarray}
W.x &=& \left(-\int_0^t x(t')\dot{f}(t') dt'\right)x(t) \nn\\
&=& \left[-\int_0^t\left(x_0e^{-kt'/\gamma}+\frac{1}{\gamma}\int_0^{t'}e^{-k(t'-t'')/\gamma}(f(t'')+\xi(t''))dt''\right)\dot{f}(t')dt'\right] \nn\\
&& ~~~~~~~~~~~~~~~\times\left[x_0e^{-kt/\gamma}+\frac{1}{\gamma}\int_0^te^{-k(t-t_1)/\gamma}(f(t_1)+\xi(t_1))dt_1\right]\nn
\end{eqnarray}

\begin{eqnarray}
\therefore \la W.x\ra &=& -\int_0^t \la x_0^2 \ra e^{-k(t+t')/\gamma}\dot{f}(t')dt'\nn\\
&&-\frac{1}{\gamma^2}\int_0^t dt' \dot{f}(t')\int_0^{t'}dt''e^{-k(t'-t'')/\gamma}\int_0^t dt_1[f(t'')f(t_1)+\la\xi(t'')\xi(t_1)\ra]e^{-k(t-t_1)/\gamma}\nn\\
&=& -\frac{T}{k}\int_0^t e^{-k(t+t')/\gamma}\dot{f}(t')dt' \nn\\
&& -\frac{1}{\gamma^2}\int_0^t dt' \dot{f}(t')\int_0^{t'}dt''e^{-k(t'-t'')/\gamma}\int_0^t dt_1[f(t'')f(t_1)+2T\gamma\delta(t-t')]e^{-k(t-t_1)/\gamma},\nn\\
\label{e12}
\end{eqnarray}
\end{widetext}

where we have used the fact that $\frac{1}{2}k\la x_0\ra^2 = \frac{1}{2} T$, and $\la \xi(t)\xi(t')=2T\delta(t-t')$. Also, $x_0$ and $\xi(t)$ are uncorrelated.

From (\ref{e11}) and (\ref{e12}),

\begin{eqnarray}
&&\la W(t)x(t)\ra - \la W(t)\ra\la x(t)\ra\nn\\
&=& -(T/k)\int_0^t e^{-k(t+t')/\gamma}\dot{f}(t')dt' \nn\\
&& -(2T/\gamma)\int_0^t dt'\dot{f}(t')\int_0^{t'}e^{-k(t'-t'')/\gamma}e^{-k(t-t'')/\gamma}dt'', \nn\\
 &=& -(T/k)e^{-kt/\gamma}\int_0^te^{-kt'/\gamma}\dot{f}(t')dt' \nn\\
&& -(2T/\gamma)e^{-kt/\gamma}\int_0^t dt'\dot{f}(t')e^{-kt'/\gamma}\int_0^{t'}e^{2kt''/\gamma}dt''.
\end{eqnarray}

Finally, one obtains

\beq
\la W(t)x(t)\ra - \la W(t)\ra\la x(t)\ra = -\frac{T}{k}e^{-kt/\gamma}\int_0^t dt'\dot{f}(t')e^{kt'/\gamma}dt'.
\label{eqcorr}
\eeq

On integrating by parts, the integral on the RHS becomes

\bea
\left[e^{kt'/\gamma}f(t')\right]_0^t &-& \int_0^t\frac{k}{\gamma}e^{-kt'/\gamma}f(t')\nn\\
&=& e^{kt/\gamma}f(t)-\frac{k}{\gamma}\int_0^t e^{kt'/\gamma} f(t')dt'. \nn
\label{e14}
\eea

Using this, equation (\ref{eqcorr}) reduces to

\beq
\la W(t)x(t)\ra - \la W(t)\ra\la x(t)\ra = \frac{T}{k}[k\la x(t)\ra - f(t)].
\label{cc}
\eeq

Finally, from (\ref{cc}) and (\ref{eqvar1}), we get
\beq
\sigma^2 = \frac{T}{T}[2\la W\ra - k\la x\ra^2 + 2\la x\ra f]=2\la \Delta s_{tot}\ra.
\eeq

\section{Calculation of the Fourier transform of $P(\Delta s_{tot},t)$}
\label{appB}

\bea
&&\widehat{P}(R,t) \equiv \int_{-\infty}^{\infty}d\Delta s_{tot}e^{iR\Delta s_{tot}}P(\Delta s_{tot},t)\nn\\
&=& \int_{-\infty}^{\infty}dx ~dx_0P(x_0,x,t)\exp\left[iR\left(\frac{\alpha}{2}x_0^2+\frac{\beta}{2}x^2+\kappa\right)\right]\nn\\
&=& e^{iR\kappa}\int_{-\infty}^{\infty}dx ~dx_0P(x_0,x,t)\exp\left[iR\left(\frac{\alpha}{2}x_0^2+\frac{\beta}{2}x^2\right)\right].\nn\\
\label{eq51}
\eea

The factor $\exp\left[iR\left(\frac{\alpha}{2}x_0^2+\frac{\beta}{2}x^2\right)\right]$ in (\ref{eq51}) can be written as

\beq
\exp\left[iR\left(\frac{\alpha}{2}x_0^2+\frac{\beta}{2}x^2\right)\right]=e^{\frac{1}{2}iR{\bf a}^\dagger .{\bf B.a}},
\eeq

with

\beq
{\bf B}\equiv\left(\begin{array}{cc}\alpha & 0\\ 0 & \beta \end{array}\right).
\eeq

\bea
\therefore \widehat{P}(R,t) &=& \frac{e^{iR\kappa}}{2\pi\sqrt{\det {\bf A}}}\int_{-\infty}^{\infty}d{\bf a} ~e^{-\frac{1}{2}{\bf a^\dagger. A^{-1}.a}+i\frac{R}{2}{\bf a^\dagger.B.a}}\nn\\
&=& \frac{e^{iR\kappa}}{2\pi\sqrt{\det {\bf A}}}\int_{-\infty}^{\infty}d{\bf a} ~ e^{-\frac{1}{2}{\bf a}^\dagger.({\bf A}^{-1}-iR{\bf B}).{\bf a}}\nn\\
&=& \frac{e^{iR\kappa}}{2\pi\sqrt{\det {\bf A}}}\int_{-\infty}^{\infty}d{\bf a} ~ e^{-\frac{1}{2}{\bf a^\dagger. A^{-1}.}({\bf I}-iR{\bf A.B}){\bf .a}}\nn\\
&=& \frac{e^{iR\kappa}}{2\pi\sqrt{\det {\bf A}}}\frac{2\pi}{\sqrt{\det ({\bf A}^{-1})\det({\bf I}-iR{\bf A.B})}}\nn\\
&=& \frac{e^{iR\kappa}}{\sqrt{\det({\bf I}-iR{\bf A.B})}}.
\label{eq54}
\eea

which is equation (\ref{eqPhat}).

The determinant $\det({\bf I}-iR{\bf A.B})$ is given by

\begin{eqnarray}
&&\det (I-iR{\bf A.B}) = \frac{T-iR(\sigma_x^2-T)}{k\la x^2\ra}\nn\\
&&+ \left(\frac{\sigma_x^2-T}{k\la x^2\ra}\right)e^{-2kt/\gamma}\left[(1+iR)^2+R(i-R) \right.\nn\\
&&\left.\left\{\left(\frac{\sigma_x^2-T}{T}\right)e^{-2kt/\gamma}-\frac{\sigma_x^2}{T}\right\}\right].
\end{eqnarray}



\section{Proof of  {\boldmath $\la\Delta F(\tau)\ra \ge \Delta F$}  for harmonic oscillator}

\label{appC}

In this appendix, our motivation is to evaluate $\la\Delta F(\tau)\ra$ and show that $\la\Delta F(\tau)\ra \ge \Delta F$.

Let us consider the potential 

\beq
U(x,\tau)=\frac{1}{2}kx^2-xf(\tau),
\eeq

where $f(t)$ is an arbitrary protocol. 
The protocol $\lambda(t) = f(t)$ is assumed to be equal to zero at time $t=0$. Thus, $\lambda(0)=0$. After time $\tau$, $\lambda(\tau)=f(\tau)$. The equilibrium free energy at parameter corresponding to $t=0$, is $F_A = T\ln\left(\sqrt{\frac{k}{2\pi T}}\right)$. 
The equilibrium free energy corresponding to the final value of the protocol is

\beq
F_B =  T\ln\left(\sqrt{\frac{k}{2\pi T}}\right)-\frac{f^2}{2k}.
\eeq

Here,

\beq
\Delta F = F_B-F_A = -\frac{f^2}{2k}.
\eeq

The initial probability density of the particle position is 

\beq
P(x_0)=\sqrt{\frac{k}{2\pi T}}\exp\left(\frac{-kx_0^2}{2T}\right).
\eeq

The final time-evolved solution for $P(x,\tau)$ is

\beq
P(x,\tau)=\sqrt{\frac{k}{2\pi T}}\exp\left(\frac{-k(x-\la x\ra)^2}{2T}\right).
\eeq

where $\la x(\tau)\ra$ is obtained from equation (\ref{eq5}) on replacing $t$ by $\tau$.
Thus,

\bea
\la \Delta F(\tau)\ra-\Delta F &=& \frac{1}{2}k\la x^2\ra -\la x\ra f -\frac{T}{2}  + \frac{f^2}{2k} \nn\\
&=& \frac{1}{2}k\left(\frac{T}{k}+\la x\ra^2\right) - \la x\ra f -\frac{T}{2} + \frac{f^2}{2k}\nn\\
&=&\frac{1}{2}k\left(\la x\ra^2-2\la x\ra\frac{f}{k}+\frac{f^2}{k^2}\right)\nn\\
&=& \frac{1}{2}k\left(\la x\ra-\frac{f}{k}\right)^2 \ge 0.
\eea

When $f(t)=A\sin\omega t$, the instantaneous change in free energy is given by

\begin{eqnarray}
&&\Delta F(t) = \frac{1}{2}k\la x(t)\ra^2 - \la x(t)\ra f(t) \nn\\
&=& \frac{A^2e^{-kt/\gamma}\sin\omega t}{k^2+\gamma^2\omega^2}\left[\gamma\omega + e^{kt/\gamma}(-\gamma\omega\cos\omega t+k\sin\omega t)\right]\nn\\
&+&\frac{kA^2e^{-2kt/\gamma}}{2(k^2+\gamma^2\omega^2)^2}\left[\gamma\omega + e^{kt/\gamma}(-\gamma\omega\cos\omega t+k\sin\omega t)\right]^2.\nn\\
\end{eqnarray}

and change in equilibrium free energy is given by

\beq
\Delta F = \frac{A^2\sin^2\omega t}{2k}.
\eeq

For a protocol of time interval between $t=0$ to $t=\tau=2\pi/\omega$, we get

\begin{widetext}
\begin{equation}
\la \Delta F(\tau)\ra = -\frac{A^2\left[k^3+\left(2-e^{-k\pi/\gamma\omega}\right)k\gamma^2\omega^2+2e^{-k\pi/2\gamma\omega}\gamma^3\omega^3\right]}{2(k^2+\gamma^2\omega^2)^2}; ~~~~
\la\Delta F\ra = -\frac{A^2}{2k}.
\end{equation}
\end{widetext}



\end{document}